\documentclass[american]{elsarticle}
\usepackage{babel}
\usepackage[utf8]{inputenx}
\usepackage{csquotes}
\usepackage{amsmath}
\usepackage{graphicx}
\usepackage{mathtools}
\usepackage{array}
\usepackage{url}
\usepackage{tikz}
\usetikzlibrary{arrows.meta}
\usetikzlibrary{positioning}
\usepackage{tikz-dimline}
\pgfplotsset{compat=1.14}
\usepackage{xcolor}
\usepackage{siunitx}
\colorlet{lightgray}{black!15}
\usepackage{environ}
\usepackage{booktabs}
\usepackage{tabu}
\usepackage[inline,shortlabels]{enumitem}
\usepackage{IEEEtrantools}
\usepackage[caption=false,font=footnotesize]{subfig}
\usepackage{microtype} 
\usepackage[hidelinks]{hyperref}

\makeatletter
\newsavebox{\measure@tikzpicture}
\NewEnviron{scaletikzpicturetowidth}[1]{%
  \def\tikz@width{#1}%
  \def\tikzscale{1}\begin{lrbox}{\measure@tikzpicture}%
  \BODY{}
  \end{lrbox}%
  \pgfmathparse{#1/\wd\measure@tikzpicture}%
  \edef\tikzscale{\pgfmathresult}%
  \BODY{}
}
\makeatother

\newcommand{\esp}[1]{\mathrm{E}\left[{#1}\right]}
\newcommand{\Tany}[1]{\mathrm{T}_{\mathrm{#1}}}
\newcommand{\TLPI}{\Tany{LPI}}
\newcommand{\TON}{\Tany{ON}}
\newcommand{\Tcycle}{\Tany{cycle}}
\newcommand{\TS}{\Tany{S}}
\newcommand{\TW}{\Tany{W}}
\newcommand{\sLPI}{\sigma_{\mathrm{LPI}}}
\newcommand{\rLPI}{\rho_{\mathrm{LPI}}}

\title{Leveraging Energy Saving Capabilities of Current EEE Interfaces via
  Pre-Coalescing\tnoteref{t1,t2}}

\tnotetext[t1]{This work was supported by the European Regional Development
  Fund (ERDF) and the Galician Regional Government under agreement for funding
  the Atlantic Research Center for Information and Communication Technologies
  (\href{https://atlanttic.uvigo.es}{atlanTTic}).}

\tnotetext[t2]{This work was funded by the ``Ministerio de Economía, Industria
  y Competitividad'' through the project TEC2017-85587-R 
  of the ``Programa Estatal de Investigación, Desarrollo e Innovación
  Orientada a los Retos de la Sociedad'', (partly financed with FEDER funds).}

\author{Miguel Rodríguez-Pérez\corref{cor1}}
\ead{miguel@det.uvigo.gal}
\author{Sergio Herrería-Alonso}
\ead{sha@det.uvigo.es}
\author{Raúl F. Rodríguez-Rubio}
\ead{rrubio@det.uvigo.es}
\author{José Carlos López-Ardao}
\ead{jardao@det.uvigo.es}

\cortext[cor1]{Corresponding author}

\address{Telematics Engineering Department, EE Telecomunicación, Rúa Maxwell
  s/n, University of Vigo, 36310 Vigo, Spain.}
  
\begin{document}

\begin{abstract}
  The low power idle mode implemented by Energy Efficient Ethernet (EEE)
  allows network interfaces to save up to \SI{90}{\percent} of their nominal
  energy consumption when idling. There is an ample body of research that
  recommends the use of frame coalescing algorithms---that enter the low power
  mode as soon as there is no more traffic waiting to be sent, and delay the
  exit from this mode until there is an acceptable amount of traffic
  queued---to minimize energy usage while maintaining an acceptable
  performance. However, EEE capable hardware from several manufactures delays
  the entrance to the low power mode for a considerable amount of time
  (hysteresis). In this paper we augment existing EEE energy models to account
  for the hysteresis delay and show that, using the configuration ranges
  provided by manufacturers, most existing EEE networking devices are unable
  to obtain significant energy savings. To improve their energy efficiency, we
  propose to implement frame coalescing directly at traffic sources, before
  reaching the network interface. We also derive the optimum coalescing
  parameters to obtain a given target energy consumption at the EEE device
  when its configuration parameters are known in advance.
  
  \vspace{1ex}
  \copyright{} 2020 Elsevier Ltd. This manuscript version is made available under the \href{http://creativecommons.org/licenses/by-nc-nd/4.0}{CC-BY-NC-ND 4.0 license}\\DOI: \href{http://dx.doi.org/10.1016/j.jnca.2020.102734}{10.1016/j.jnca.2020.102734}.
\end{abstract}


\maketitle

\section{Introduction}
\label{sec:introduction}

Energy Efficient Ethernet (EEE) interfaces are already widespread, as they
have been around us for the last ten
years~\cite{802.3az,christensen10:_the_road_to_eee}. These interfaces are able
to avoid wasting precious energy when they are not transmitting data. For
this, they implement one or more low power modes~\cite{802.3az,IEEE802.3bj}
that only spend a fraction of the nominal power usage, but that cannot be
employed during normal data transmission. However, going in and out from these
low power modes is not free, as it takes some time and energy that are
otherwise unavailable for doing useful work.

The compromise between energy efficiency and the time spent transitioning to
and from low power idle (LPI) modes has been the subject of many in-depth
analyses, as it has direct consequences in two very important metrics: energy
consumption and traffic delay. A simple, yet efficient, way to improve energy
efficiency consists in amortizing the transitions among several frames.
Instead of exiting LPI as soon as a new frame arrives, it is more efficient to
wait until a larger set of frames are waiting for transmission and then
perform a single transition back to the active state for the whole set. This
coalescing approach~\cite{reviriego10:_burst_tx_eee} clearly trades some delay
for greater energy
savings~\cite{Kim2013,herreria11:_power_savin_model_burst_trans}. Many
coalescing proposals employ the number of queued frames as the condition to
exit LPI, but time-based coalescers, that use the time since the first frame
arrival while in LPI, can also be configured to obtain identical
results~\cite{Pan2017,herreria19:_dynam_eee_coales}.

A complementary approach to avoid excessive transitions consists in delaying
the entrance into the LPI mode when the transmission queue is drained. In this
case, the EEE interface waits for new immediate arrivals during a small
additional hysteresis time before entering the LPI mode. Thus, if the time
until the next frame arrival is shorter than this hysteresis time, traffic
does not get unnecessarily delayed and some energy is saved, since entering
the LPI mode is not efficient enough to compensate for the energy consumed
during the state transitions. However, if the frame arrives just after the
hysteresis has ended, traffic gets delayed and the energy employed waiting
during the hysteresis time becomes wasted. In fact, it has been proved that
for low traffic loads, adding hysteresis results in greater energy
usage~\cite{Herreria-Alonso2018b}.

Most hardware manufacturers have included EEE low power modes in their
products. Many of them have chosen to implement both a hysteresis delay and a
time-based frame coalescer to improve efficiency.\footnote{Regretfully, not
  all manufacturers provide details about the actual implementation
  characteristics of their EEE mode. Even more, some simply let the
  administrator enable or disable the EEE mode, but not to tune its
  parameters.} However, their tuning of these parameters, and the available
configuration range, is too much conservative, as we will prove later. Minimum
values higher than \SI{20}{\us} are normal for the hysteresis length of
\SI{10e9}{\bit\per\s} interfaces, even reaching hundreds of microseconds.
Regarding the coalescing timer, when it is available, its maximum configurable
value is sometimes too short to compensate for high hysteresis times.
\begin{table}
  \centering
  \caption{Available Configuration Options of Several Providers of EEE Capable
    Devices}
  \label{tab:manufactures-review}
  \begin{tabu}{XSc}
    \toprule
    \textbf{Manufacturer}&\textbf{Hysteresis (\si{\us})}&\textbf{Coalescing (\si{\us})}\\\midrule
    Cisco Nexus 7000~\cite{cisco7000}&\numlist{20;600}&6\\
    Dell EMC N-Series~\cite{dell1524p}&\numrange{600}{4294967295}&\numrange{0}{65535}\\
    QLogic bnx2~\cite{mintz-qlogic-eee-patch}&\numrange{256}{1048575}&?\\
    Intel
    X550~\cite{intel_ether_contr_x550_datas} \& X710~\cite{intel_ether_contr_x710_xxv71_xl710_datas}&\numrange{1}{63}&?\\
    D-Link DGS-1100–16~\cite{sivaraman14:_energ_effic_ether}&\num{0}&\num{0}\\
    Level-One GEU-0820~\cite{sivaraman14:_energ_effic_ether}&\num{0}&\num{0}\\
    SMC GS801~\cite{sivaraman14:_energ_effic_ether}&\num{0}&\num{0}\\
    \bottomrule
  \end{tabu}
\end{table}
In Table~\ref{tab:manufactures-review} we provide a small sample of the
available configuration ranges for both the hysteresis and coalescing timers
of some popular EEE devices. Note that the last three devices feature
\SI{1e9}{\bit\per\s} ports, but we added them for completeness, as they have
been thoroughly analyzed in~\cite{sivaraman14:_energ_effic_ether}.

This paper analyzes the energy efficiency limits of actual networking devices.
For this, we contribute a new model for EEE interfaces with hysteresis. The
model is completed to include the joint effects of hysteresis and time-based
coalescing. The model shows that for most practical loads, and with the
available values for hysteresis and coalescing delays, the LPI mode in some
EEE hardware is not sufficient to reduce energy consumption. Our second
contribution is a proposal to deploy time-based coalescers before the Network
Interface Card (NIC) so as to overcome the limitations of networking
equipment. This technique helps to overcome some of the drawbacks of devices
that employ high hysteresis values and can also augment the saving of devices
that lack any coalescing capabilities. We provide an analytic model for this
technique and also solve its proper tuning to obtain any desired approximation
to the optimal efficiency.

The rest of this paper is organized as follows. Section~\ref{sec:related-work}
summarizes previous work dealing with EEE\@. In
Section~\ref{sec:hardware-model} we present a power model for the behavior of
actual hardware and provide early estimations of the expected energy savings.
The pre-coalescing solution is shown and modeled in
Section~\ref{sec:source-based-coals}. We test the behavior of our solution in
Section~\ref{sec:experimental-results} and finally, we present our conclusions
in Section~\ref{sec:conclusions}.

\section{Related Work}
\label{sec:related-work}

The lack of any prescribed governing algorithm for the LPI mode of EEE
in~\cite{802.3az} led to the development of many competing proposals. The
simplest ones arrived soon and consisted in simply entering LPI as soon as
there was no traffic left to be transmitted to then resume normal operation as
soon as there was newly available traffic. These proposals are collectively
known as \emph{frame transmission} algorithms and their energy and delay
performance models are already well
known~\cite{larrabeiti11:_towar_gb_ether,ajmone11,bolla14,RodriguezPerez2017}.
These energy consumption models show that frame transmission causes a high
number of transitions to and from LPI that severely hinders its performance,
thus rendering it a poor choice for most traffic patterns and
loads---recall that during these transitions the network interface draws about
the same power than an active one.

The solution to this problem is to amortize these transitions among several
frames. For this, the interface is not immediately woken up when there is
newly available traffic, instead it is woken up when some more traffic has
been accumulated. Some early works that propose this \emph{frame coalescing}
(or \emph{burst transmission} as it is also known)
are~\cite{christensen10:_the_road_to_eee,reviriego10:_burst_tx_eee}. There
exist two main approaches for deciding when to wake up the interface.
Size-based coalescers put an upper bound to the queue length while in LPI and
resume normal operations once this threshold is reached.\footnote{To avoid
  excessively long waiting times, a safeguard in the form of a maximum
  sleeping time is usually implemented.} On the other hand, time-based
coalescers use the time spent in LPI, or since the first arrival while in LPI,
to decide when to restart transmitting traffic. There is ample literature
modeling the energy
consumption~\cite{herreria11:_power_savin_model_burst_trans} or even the
energy-delay tradeoffs, either just for size-based~\cite{Mostowfi2012}
coalescers, for time-based~\cite{Akar2013} ones or even for the general
case~\cite{Herreria-Alonso2012c,Chatzipapas2016,Pan2017,Kim2013,Meng2017}.

These models have been later employed to tune the frame coalescing algorithm,
i.e., deriving the optimum upper bound, so as to obtain a given energy
efficiency~\cite{Pan2017} or to meet a target average delay by limiting the
maximum time in LPI~\cite{Chatzipapas2016,Chatzipapas2016a}.

To the best of our knowledge, the introduction of some hysteresis time before
entering LPI has not yet been considered in the research literature in the
context of EEE, although it has been studied in analogous scenarios. Such is
the case of cellular base stations, in order to reduce the number of
transitions between different operating modes~\cite{Niu2015,Guo2016}. However,
as shown in~\cite{Herreria-Alonso2018b}, adding hysteresis can sometimes lead
to increased energy consumption and it is largely unneeded when using a
properly tuned frame coalescing algorithm.

\section{Hardware Model}
\label{sec:hardware-model}

\begin{figure}
  \centering
  \begin{scaletikzpicturetowidth}{\columnwidth}
  \begin{tikzpicture}[scale=\tikzscale]

    \begin{scope}
      \clip (0, 0) rectangle (8.5, 1);
      \foreach \i in {0, .8, 1.6, 7.2, 8} { \filldraw[fill=darkgray] (\i,0)
        rectangle (\i+1,.7);
      }
      \filldraw[fill=darkgray] (3, 0) rectangle (4.5, .7);
    \end{scope}

    \fill[fill=lightgray] (5.6, 0) rectangle (7, .7);
    
    \draw[->] (1.8,1.2) -- (1.5,0) node[below] {$a_i$} ;
    \draw[->] (3.3,1.2) -- (3,0) node[below] {$a_{i+1}$};

    \draw[->] (6.3,1.2) -- (6,0) node[below] {$a_{i+2}$};
    \draw[->] (7,1.2) -- (6.7,0) node[below] {$a_{i+3}$};

    \draw[loosely dotted] (7.5,1) -- (8,1);
    \draw[loosely dotted] (.5,1) -- (1.1,1);
    
    \draw[-{Stealth[length=3mm]}] (0,0) -- (8.88, 0) node[pos=.95, below]
    {$t$};

    \footnotesize
    \draw[-{Stealth}|] (0,1.5) -- node[above] {ON} (2.6,1.5) node (ON0) {};
    \draw[|{Stealth}-{Stealth}|] (3.0,1.5) -- node[above] {ON} (4.5,1.5) node (ON1) {};
    \draw[|{Stealth}-] (7.2,1.5) -- node[above] {ON} (8.8,1.5);
    \draw[|{Stealth}-{Stealth}|] (5.6, 1.5) -- node[above] {LPI} (7, 1.5) node
    (LPI) {};
    \draw[|{Stealth}-{Stealth}|] (2.6,1.5) -- node[above] {HYST} (3.0,1.5)
    node (HYST1) {};
    \draw[|{Stealth}-{Stealth}|] (4.5,1.5) -- node[above] {HYST} (5.4,1.5)
    node (HYST2) {};   
    \node [right=0 of HYST2,rotate=90] {SLEEP};
    \node [right=0 of LPI,rotate=90] {WAKE};

    \dimline[extension start length=1, extension end length=1] {(2.6,1)} {(3.5,1)} {$h^\ast$};
    \dimline[extension start length=1, extension end length=1.1] {(4.5,1)} {(5.4,1)} {$h^\ast$};
    \dimline[extension start length=1, extension end length=1] {(6,1)} {(7,1)} {$d$};    
  \end{tikzpicture}
  \end{scaletikzpicturetowidth}
  \caption{EEE operation with hysteresis and delay times.}
  \label{fig:modus-operandi}
\end{figure}
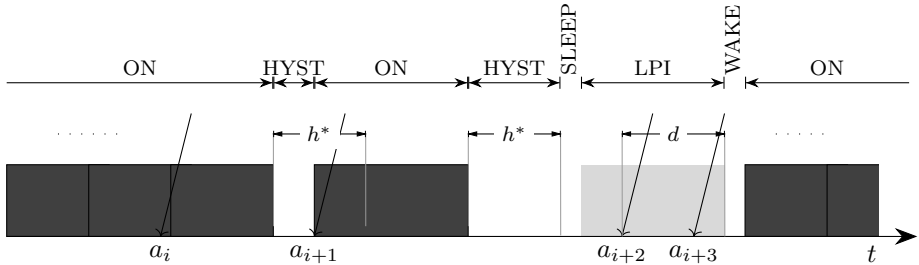

As it was already stated in the previous sections, currently deployed EEE
interfaces use a time-based coalescing algorithm with hysteresis to drive the
LPI mode, as represented in Fig.~\ref{fig:modus-operandi}.
That is, after the transmission queue becomes depleted, most EEE interfaces
wait for the queue to remain empty for at least $h^\ast$ seconds before
entering the LPI mode, with $h^\ast$ being the \emph{hysteresis} time.
Conversely, normal operation resumes $d$ seconds after the first frame arrival
in LPI\@. This coalescing time $d$ is simply called \emph{delay}. As
aforementioned in Section~\ref{sec:related-work} the performance of EEE with
hysteresis has not been studied yet in the literature, so we will provide an
extended EEE energy usage model in this section taking it into account.

The purpose of the hysteresis time is to avoid entering LPI when the load is
high, i.e., when frames arrive too close to each other. When this happens,
entering and exiting LPI increases both delay and energy consumption, as the
interface consumes about the same amount of energy during transitions as
during regular transmissions. Hysteresis works on the idea that if the
expected time in LPI is much smaller than the sum of the \emph{sleep} and
\emph{wake up} transition times, the energy savings are probably not worth the
added delay. Once the interface is in LPI, it is important to keep it in that
state as long as possible to amortize the transition delays, but with a
controlled latency. That is the objective of the \emph{delay} value. In
essence, it plays an analogous role to the queue threshold in EEE size-based
coalescing
algorithms~\cite{reviriego10:_burst_tx_eee,Kim2013,herreria11:_power_savin_model_burst_trans}.

In the following subsections, we will derive a power saving model for EEE
interfaces with both a time-based coalescer, i.e., $d>0$, and hysteresis, i.e., $h^\ast>0$. 

\subsection{Energy Model Considering Hysteresis}
\label{sec:energy-model-with}

As shown in~\cite{Herreria-Alonso2012c},
the average normalized energy consumption of an EEE interface can be calculated as
\begin{equation}
  \label{eq:energy-consumption}
  \sigma = 1-(1-\sLPI)\rLPI,
\end{equation}
where $\sLPI$ is the normalized energy consumption of the LPI mode and
\begin{equation}
  \label{eq:rho-off-definition}
  \rho_{\mathrm{LPI}}=\frac{\esp\TLPI}{\esp\Tcycle}
\end{equation}
is the normalized sleeping time, that is, the percentage of time the interface
stays in LPI, as $\TLPI$ is the duration of the LPI sub-period and $\Tcycle$
the time between two consecutive transitions from LPI to the normal operating
mode.

Clearly, in a given cycle there can only be a single LPI period and,
consequently, one transition to and from it. However, there can be several
\emph{ON} periods, as a single arrival during the hysteresis time prevents the
interface from entering the low power mode. So
\begin{equation}
  \label{eq:tcycle}
  \esp\Tcycle = \TS + \esp\TLPI + \TW + \esp n(\esp\TON +\esp h),
\end{equation}
with $n$ the number of times the interface fails to enter LPI while in
hysteresis, $h \leq h^\ast$ a random variable representing the actual length
of a hysteresis interval, and $\TON$ the length of an ON interval. $\TS$ and
$\TW$ are the transition times from the ON state to LPI and from LPI back to
ON\@. They only depend on the interface characteristics, so we treat them as
known constants.

If we assume a conservative service at the interface, it must hold that
\begin{equation}
  \label{eq:rho-new-cycle}
  \rho = \frac{\esp n\esp\TON}{\esp\Tcycle},
\end{equation}
with $\rho$ being the load factor, and, from~\eqref{eq:tcycle}
and~\eqref{eq:rho-new-cycle}, it immediately follows that
\begin{equation}
  \label{eq:rho-off-analytic}
  \rho_{\mathrm{LPI}} = (1-\rho)\frac{\esp\TLPI}{\esp\TLPI + \esp n\esp h + \TS+\TW}.
\end{equation}

The average length of the hysteresis interval is simply
\begin{equation}
  \label{eq:esp-h-general}
  \esp h = \lambda \int_0^\infty \min(t, h^*) f_e(t)\,\mathrm{d}t,
\end{equation}
where $f_e(t)$ is the density function of the time between the
end of the ON state and the next frame arrival.
Additionally, if arrivals are independent, $\esp n$ is just the average value
of a geometric distribution parameterized by the probability that there are no
arrivals during the hysteresis interval.

In the specific case of a Poisson
process with interarrival times $I$ and arrival rate $\lambda$ it holds that
\begin{equation}
  \label{eq:esp-n-poisson}
  \esp n = \frac{1}{\mathrm{P}(I>h^\ast)} = \mathrm{e}^{\lambda h^\ast}.
\end{equation}
and, because of the PASTA property, $f_e(t)$ is the density function of an
exponential distribution with parameter $\lambda$, so
\begin{equation}
  \label{eq:esp-h}
  \esp h = \lambda \int_ 0^{h^\ast}t\mathrm{e}^{-\lambda t}\,\mathrm{d}t + \lambda
  h^\ast \int_{h^\ast}^{\infty} \mathrm{e}^{-\lambda t}\,\mathrm{d}t =
  \frac{(1-\mathrm{e}^{-\lambda h^\ast})}\lambda.
\end{equation}

\subsection{Model for Time-Based Coalescers with Hysteresis}
\label{sec:model-time-based}

We can extend the previous model to take into account the effect of the delay
timer. We need to obtain the average time from the end of the transition to
LPI until $d$ seconds after the arrival of the first frame since the end of
hysteresis. Particularizing directly for a Poisson arrival process, we get

\begin{IEEEeqnarray}{rCl}
  \label{eq:esp-tlpi}
  \esp{\TLPI} &=& \lambda \int_{{(\TS-d)}^+}^{\infty} (t+d-\TS)\,\mathrm{e}^{-\lambda
    t}\,\mathrm{d}t\nonumber\\
  &=&
    \begin{cases}
      \lambda^{-1} + d-\TS,& \text{ for } d > \TS,\\
      \frac{\mathrm{e}^{-\lambda\left(\TS-d\right)}}\lambda, & \text{ in every other case,}\\
    \end{cases}
\end{IEEEeqnarray}
where ${(x)}^+ = \max\{x, 0\}$.

\subsection{Model Results for Real Hardware}
\label{sec:model-results-real}

\begin{figure}
  \centering
  \includegraphics[width=\columnwidth]{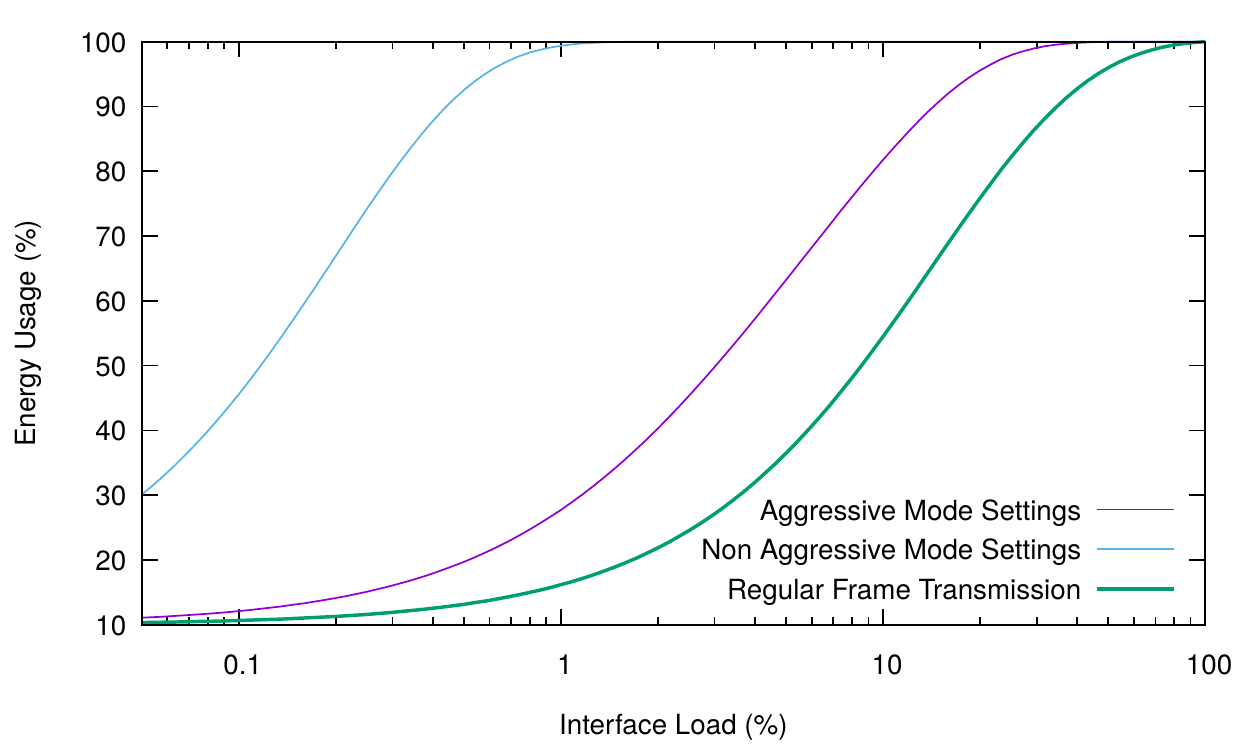}
  \caption{Expected energy usage for different values of $h^\ast$ and $d$.}
  \label{fig:expected-savings}
\end{figure}
With the previous model, we can take a closer look at the expected energy
performance of some representative hardware. According to configuration
manuals from several manufactures, e.g.,~\cite{cisco7000,dell1524p}, the delay
parameter takes a default value in the order of \SIrange{6}{8}{\us}, while the
minimum allowable hysteresis value can get as high as \SI{600}{\us} for some
of them (see Table~\ref{tab:manufactures-review}).
In Fig.~\ref{fig:expected-savings} we compare the expected energy usage for an
extremely ample range of traffic loads and different settings.

We have employed two combinations of the hysteresis and delay parameters
called \emph{aggressive mode} and \emph{non aggressive mode} as shown in
Table~\ref{tab:switch-parameters}.
We use these values as representative of both extremes of available minimum
hysteresis values, and use $d=\SI{6}{\us}$ as it is the default
for~\cite{cisco7000}. We ignored interfaces such
as~\cite{intel_ether_contr_x710_xxv71_xl710_datas} as, for practical purposes,
their hysteresis can be deactivated.
\begin{table}
  \centering
  \caption{Configuration Parameters for the EEE Interface}
  \label{tab:switch-parameters}
  \begin{tabu}{rSS}\toprule
    \textbf{Interface Configuration}&\textbf{Hysteresis (\si{\us})}&\textbf{Delay
      (\si{\us})}\\\midrule
    Aggressive & 20 & 6\\
    Non Aggressive & 600 & 6\\\bottomrule
  \end{tabu}
\end{table}

Note that the non aggressive mode settings only make sense for loads well
below \SI{1}{\percent} and that even for the so-called aggressive settings we
get much worse results than with the naive frame transmission algorithm with
no coalescing whatsoever. Furthermore, the per frame maximum additional
latency is smaller for frame transmission, as it adds a maximum delay of
$\TS+\TW = \SI{7.36}{\us}$,\footnote{Using $\TS = \SI{2.88}{\us}$ and 
$\TW = \SI{4.48}{\us}$ according to~\cite{802.3az}.} while typical interface 
settings are adding an additional delay of at least
$\min\{d\} + \TW =\SI{6}{\us}+\SI{4.48}{\us}=\SI{10.48}{\us}$.

\section{Pre-Coalescing}
\label{sec:source-based-coals}

Such disappointing results may be overcome either by increasing coalescing
delay if it is configurable or by performing frame coalescing \emph{before}
the traffic arrives at the EEE network interface (\emph{pre-coalescing}). If
the EEE NIC resides in the same hosts, this can be done by a traffic shaper
placed just before the NIC\@. The technique can also be employed if the high
hysteresis EEE NIC belongs to a switch. If its traffic is dominated by a
single contributor, the latter can pre-coalesce the traffic before sending it
to the switch. In both cases, the idea is to create artificial gaps between
trains of coalesced frames (bunches) that let the interface become idle
despite the hysteresis. This should result in an increase in the normalized
average sleeping time. Pre-coalescing, when compared to traditional frame
coalescing, has the extra advantage that it adds less latency for the same
energy savings, as we will prove later.

The pre-coalescing algorithm works as follows. When an idling terminal has a
new frame ready for transmission, it waits for $B$ seconds before actually
delivering the traffic to the network interface. Then, it dispatches
frames at link rate until the queue depletes. Finally, when the coalescer
delivers the last queued frame, it waits for the next frame arrival in order
to form a new bunch.

\subsection{Pre-Coalescer Energy Model}
\label{sec:bunching-model}

We now build an energy model of this frame pre-coalescer. We will assume for
simplicity that the coalescer receives Poisson traffic with a general, albeit
independent, frame size distribution. We are interested in the energy
consumption of the network interface, but we cannot rely on the previous
analysis as the traffic coming out of the coalescer shall not follow a Poisson
process.

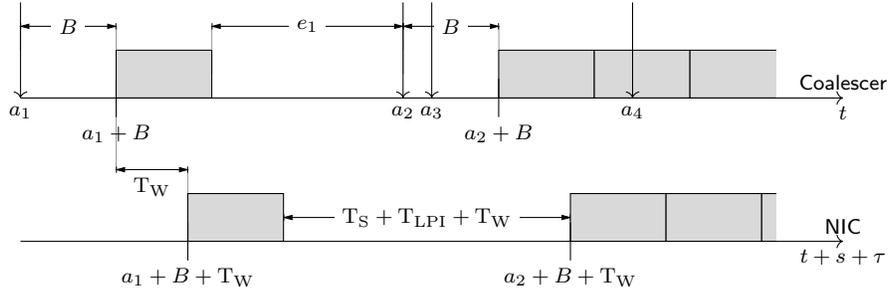
\begin{figure}
  \centering
  \begin{scaletikzpicturetowidth}{\columnwidth}
  \begin{tikzpicture}[scale=\tikzscale]
    \sffamily
    \footnotesize

    \dimline[extension start length=.25, extension end length=.25] {(2, .75)}
    {(4, .75)} {$e_1$};
    \dimline[extension start length=.25, extension end length=.25] {(0, .75)}
    {(1, .75)} {$B$};
    \dimline[extension start length=.25, extension end length=.25] {(4, .75)} {(5, .75)} {$B$};
    \dimline[extension start length=0, extension end length=0] {(2.75, -1.25)}
    {(5.75, -1.25)} {$\TS+\TLPI+\TW$};
    \dimline[label style={below}, extension start length=-1, extension end length=1] {(1, -.75)} {(1.75, -.75)} {$\TW$};

    \begin{scope}
      \clip(0, -1.5) rectangle (7.9, .6);
      \foreach \as[count=\ai] in {1, 5, 6, 7} {
        \filldraw[fill=lightgray] (\as, .5) rectangle (\as + 1, 0);
        
        \filldraw[fill=lightgray] (\as+.75, -1) rectangle (\as + 1.75, -1.5);
      }
    \end{scope}

    \draw[->] (0, 0) -- (8.6, 0) node[below] {$t$} node[above] {Coalescer};
    \draw[->] (0, -1.5) -- (8.6, -1.5) node[below] {$t+s+\tau$} node[above]
    {NIC};

    \foreach \as[count=\ai] in {0, 4, 4.3, 6.4} {
      \draw[->] (\as, 1)  --  (\as, -0) node[below] {$a_\ai$};
    }
    
    \draw (1, .2) -- +(0, -.4) node [below] {$a_1 + B$};
    \draw (5, .2) -- +(0, -.4) node [below] {$a_2 + B$};

    \draw (1.75, -1.3) -- +(0, -.4) node [below] {$a_1 + B + \TW$};
    \draw (5.75, -1.3) -- +(0, -.4) node [below] {$a_2 + B + \TW$};
  \end{tikzpicture}
  \end{scaletikzpicturetowidth}
  \caption{Time diagram of the pre-coalescing procedure. For simplicity, we
    assume $B+e_i>\TS$ and $s_i = s$.}
  \label{fig:time-diagram-bunching}
\end{figure}
Fig.~\ref{fig:time-diagram-bunching} shows a time diagram of the operation
of both the coalescer and the network interface. In the
diagram, $a_i$ are the arrival times of different frames at the coalescer,
$s_i=s$ is the corresponding frame size and $e_i$ is the elapsed time since
the transmission of the $i$-th bunch and the arrival of the next
frame.\footnote{In Fig.~\ref{fig:time-diagram-bunching} we have made $s_i=s$
  for simplicity, but the result holds for the general case.} As the arrivals
form a Poisson process of parameter $\lambda$, $e_i$ is exponentially
distributed because of the PASTA property.

We can notice in Fig.~\ref{fig:time-diagram-bunching} how the operation states
at the interface mimic those of the coalescer, with a small $\TW+s+\tau$
delay, with $\tau$ being the propagation delay. When the coalescer and the
interface reside in the same host, the propagation delay is $\tau=0$.
Furthermore, it is clear that the time between two consecutive active periods at the
coalescer must comprise both EEE transitions ($\TS+\TW$) and the time spent in
the low power mode ($\TLPI$). As both the coalescer and the interface work at
the same line rate, this must be equal to $e+B$: the time elapsed since the
end of a bunch transmission until the next one starts being delivered by the
coalescer. In other words,
\begin{equation}
  \label{eq:lpi-length-buchining-1-source-impl}
  \TS + \TW + \esp{\TLPI} = \esp{e} + B = \lambda^{-1} + B,
\end{equation}
so that
\begin{equation}
  \label{eq:lpi-length-buchining-1-source}
  \esp{\TLPI} = \lambda^{-1} + B - \TS - \TW.
\end{equation}

The cycle length is also identical both at the target interface and at the
coalescer. We can take advantage of the fact that traffic arriving at the
latter belongs to a Poisson process to obtain the average cycle length:
\begin{equation}
  \label{eq:cycle-single-pre}
  \esp{\Tcycle} = \esp{e} + B + \esp{\Tany{busy}},
\end{equation}
where $\Tany{busy}$ is the length of the coalescer busy cycle. For a M/G/1
system, $\Tany{busy} = \frac{s^\prime}{1-\rho}$, with $s^\prime$ the total
work accumulated at the start of the cycle. In our case, the accumulated
traffic at the start of transmission is just $\lambda B\cdot{}s$, so
\begin{equation}
  \label{eq:cycle-single}
  \esp{\Tcycle} = \esp{e} + B + \lambda\frac{B\cdot{}s}{(1-\rho)},
\end{equation}
and therefore,
\begin{equation}
  \label{eq:rlpi-single-nohyst}
  \rLPI = (1-\rho)
  \frac{
    \lambda^{-1} + B - \TS - \TW
  }{
    (1-\rho)\lambda^{-1} + B
  }.
\end{equation}

Please note that when $e + B < \TW$, that is, the time elapsed since the end
of the transmission until the arrival of the next frame at the coalescer plus
the bunching period is less than the transition time, the interface stops
being synchronized with the coalescer and~\eqref{eq:rlpi-single-nohyst} no
longer holds. Fortunately, this case is not of practical interest. It will be
later shown that such short $B$ values produce negligible increments in energy
savings.

\subsubsection{Effects of Hysteresis and Delay}
\label{sec:effects-hyteresis}

To model the effects of hysteresis, we proceed in a similar way. As long as
$e+B > \TW+h^\ast$, with $h^\ast$ being the hysteresis configured in the
interface, it stays in sync with the coalescer. The cycle length at the
coalescer stays the same, as it is clearly unaffected by the behavior of the
downstream interface. However, the duration between two consecutive
transmissions at the interface has to accommodate the hysteresis time, so:
\begin{equation}
  \label{eq:lpi-length-buchining-1-source-hyst}
  \TS + \TW + \esp{\TLPI} + h^\ast = \esp{e} + B = \lambda^{-1} + B,
\end{equation}
and consequently,
\begin{equation}
  \label{eq:rlpi-single-hyst}
  \rLPI^{\text{hyst}} = (1-\rho)\frac{\lambda^{-1} + B - h^\ast - \TS - \TW}{\lambda^{-1}(1-\rho) + B}.
\end{equation}

The case of added delay (time-based frame coalescing) at the interface in
addition to bunching at the coalescer is more involved. The added wake up
delay causes the interface to be delayed for $d$ time units after the first
cycle. If $d > B$, the states of the interface and the coalescer do no longer
remain in sync and we cannot apply Poisson models at the latter to model the
interface. However, usually $h^\ast \gg d$, so $B > d$. In this case, the
cycle length at the coalescer obviously stays the same as in the previous
case, and also at the interface, with the exception of the first interval that
can be neglected for our analysis. Finally, the delay time is part of $\TLPI$,
so~\eqref{eq:rlpi-single-nohyst} holds for coalescing interfaces as long as
$d < B$.

\subsection{Pre-Coalescing vs. NIC Coalescing Frame Delay}
\label{sec:pre-coalescing-vs}

The pre-coalescer energy model allows us to compare the benefits of
pre-coalescing against time-based coalescing at the interface.
\begin{figure}
  \centering
  \includegraphics[width=\columnwidth]{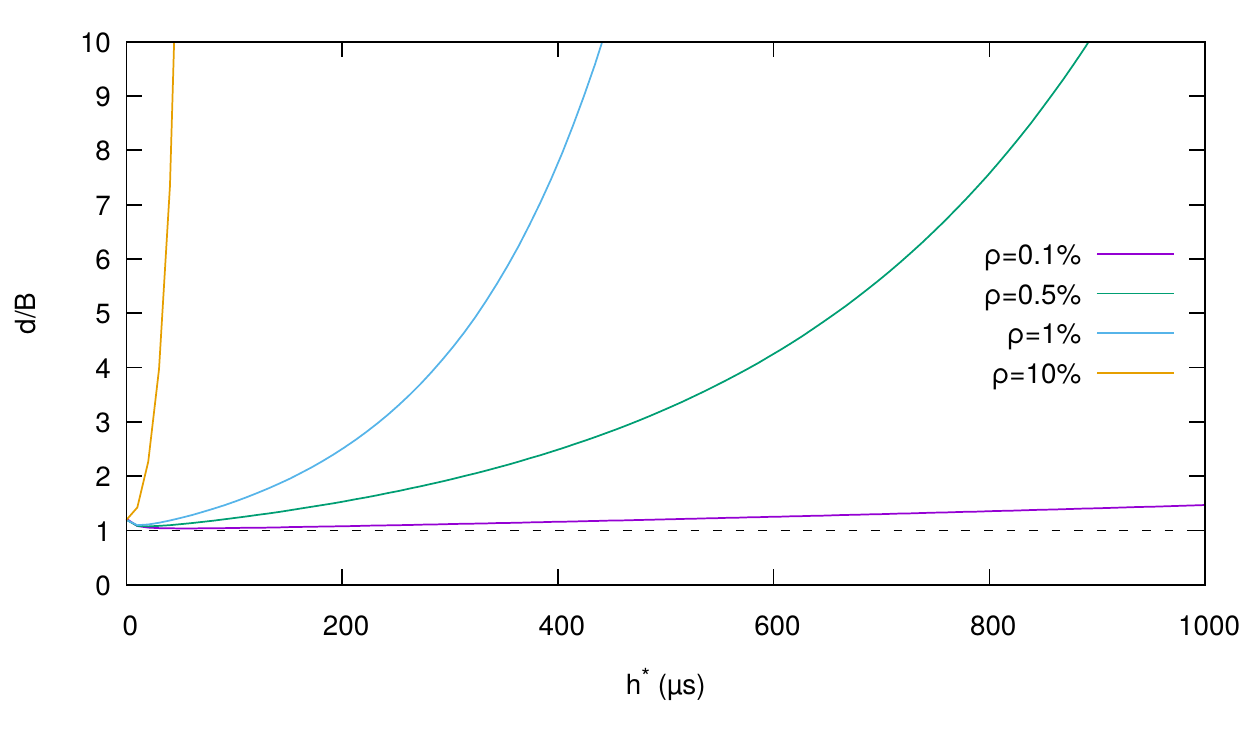}
  \caption{Relation between minimum delay ($d$) and bunch length ($B$) needed 
  to reach the same consumption for different hysteresis values ($h^\ast$) and 
  traffic loads ($\rho$).}
  \label{fig:d_vs_B}
\end{figure}
In Fig.~\ref{fig:d_vs_B} we have represented the relation between the needed
delay for a time-based coalescer ($d$) and the bunching length employed at the
source ($B$) in order to guarantee the same energy consumption, assuming a
Poisson traffic arrival process. For low traffic loads and small hysteresis
values, both parameters remain close. However, when the load is moderate or
high, pre-coalescing requires much smaller coalescing lengths. The reason is
that pre-coalescing creates gaps between bursts of frames before they get to
the interface. These gaps are, by design, longer than the configured
hysteresis, so the interface can always enter the LPI mode. However,
traditional frame coalescing at the interface only takes place \emph{after}
the interface enters LPI\@. If the average interval between frame arrivals is
smaller than the hysteresis time, as is the case with moderate loads, the
interface seldom enters LPI\@. In this case, the few times the interface does
enter LPI, it needs to stay a considerable amount of time on it to get the
same energy savings as pre-coalescing.

\subsubsection{Pre-Coalescing Algorithm Delay}
\label{sec:pre-coal-algor}

The average frame delay can be modeled considering the coalescer$+$interface
tandem as a GI/G/1 queue with added delay for the first customer in a busy
period. In fact, the coalescer is akin to a queue that waits for $B$ seconds
before serving the first job in its cycle. The network interface should add no
additional queuing delay, as both the coalescer and the interface work at the
same rate. According to~\cite{marshall1968bounds}, and considering that
waiting times and arrivals are actually uncorrelated---recall that the
pre-coalescer uses a timeout~$B$ since the first frame arrival to start
transmitting the bunch---the average waiting time of such a queue is simply
\begin{equation}
  \label{eq:av-delay-marshall-general}
  \esp W =
  \frac{\lambda(\sigma_S^2+\sigma_I^2)}{2(1-\rho)}+\frac{1-\rho}{2\lambda}+\frac{(B+\TW)^2-\esp
  {e^2}}{2(B+\TW+\esp e)},
\end{equation}
where $\sigma_S^2$ is the transmission time variance, $\sigma_I^2$ is the
variance of the arrival distribution and $e$ is the random variable representing
the length of the empty periods, that is,
the interval length since the end of a cycle until the arrival of the next
frame. In the case of a Poisson arrival process ($\sigma_I^2=1/\lambda^2$), 
recall that empty periods also follow an exponential
distribution of parameter $\lambda$ by the PASTA property, so that $\esp
e=1/\lambda$ and $\esp{e^2}=2/\lambda^2$. Substituting all these values
in~\eqref{eq:av-delay-marshall-general}, we finally arrive to
\begin{equation}
  \label{eq:av-delay-marshall-poisson}
  \esp W = \frac{1+\lambda^2\sigma_S^2}{2\lambda(1-\rho)} + \frac{1-\rho}{2\lambda}+\frac{\lambda^2(B+\TW)^2-2}{2\lambda\left(1+\lambda(B+\TW)\right)}.
\end{equation}

\subsection{Tuning the Pre-Coalescing Algorithm}
\label{sec:achi-frame-transm}

In this section we use~\eqref{eq:rlpi-single-hyst} to derive the minimum $B$
value needed for the interface to have a target energy profile. We consider two
different example targets: matching the performance of the naive frame
transmission algorithm and approximating the behavior of an ideal EEE interface.
Throughout this section we will assume that $B > \{h^\ast, d, \TW\}$.

\subsubsection{Matching Frame Transmission Consumption}
\label{sec:match-frame-coal}

It is an established fact that the frame transmission algorithm achieves less
than ideal energy savings for moderate and high traffic loads. However,
interfaces using common hysteresis values get even worse results. On the other
hand, for low traffic loads, frame transmission attains nearly-optimal
results.

The energy usage of frame transmission under the hypothesis of Poisson traffic
is well established in the
literature~\cite{larrabeiti11:_towar_gb_ether,bolla14,ajmone11,Herreria-Alonso2012c}.
From~\eqref{eq:rho-off-analytic} and~\eqref{eq:esp-tlpi}, after making
$h^\ast=0$ and $d=0$, we get that the normalized length of the frame
transmission algorithm sleeping interval is
\begin{equation}
  \label{eq:rho-lpi-frame}
  \rho_{\mathrm{LPI}}^{\mathrm f} = (1-\rho)\frac{\mathrm{e}^{-\lambda
      \TS}}{\mathrm{e}^{-\lambda \TS} + \lambda(\TS + \TW)}.
\end{equation}

Now, if we equal~\eqref{eq:rho-lpi-frame} to~\eqref{eq:rlpi-single-hyst} and
solve for $B$, after some straightforward simplifications we find that
the optimum $B$ value that matches frame transmission performance is
\begin{equation}
  \label{eq:b-for-frame}
  B^\ast_{\mathrm f} = \frac%
  {\lambda\left[%
      h^\ast+\TS+\TW+(\TS+\TW)(\lambda(h^\ast+\TS+\TW)-1)\mathrm{e}^{\lambda\TS}
    \right]%
    -\rho}%
  {(\TS+\TW)\,\lambda^2\,\mathrm{e}^{\lambda\TS}},
\end{equation}
for all but the highest rates, as $\lambda^{-1} \gg \TS$. Additionally, for
the usual hysteresis values $h \gg \{\TS, \TW\}$, and applying
the fact that, as long as the frame size stays below
\SI{1500}{bytes}, $s < \TS$ for a \SI{10e9}{\bit\per\s} interface, we get that
$B^\ast_{\mathrm f}$ can be approximated as
\begin{equation}
  \label{eq:b-for-frame-aprox}
  B^\ast_{\mathrm f} \approx h\left(
    1 + \frac{\lambda^{-1}}{\TS + \TW}
    \right).
\end{equation}

\subsubsection{Approximating Ideal EEE Interface Consumption}
\label{sec:appr-ideal-eee}

Pre-coalescing can also be used to approximate the results of an ideal EEE
interface, that is, one that stays in LPI during $1-\rho$ of the total time.
Consider $\delta$ the maximum deviation permitted from the optimum value. We
firstly equate the target consumption to~\eqref{eq:rlpi-single-hyst}:
\begin{equation}
  \label{eq:opt-b-inequality}
  1-\rho-\delta = \rLPI^{\text{hyst}} = (1-\rho)
  \frac{%
    \lambda^{-1} + B^\ast_{\mathrm i} - h^\ast - \TS - \TW
  }{%
    \lambda^{-1}(1-\rho) + B^\ast_{\mathrm i}
  }.
\end{equation}
Then, we can solve again for $B^\ast_{\mathrm i}$, the minimum $B$ value that
guarantees an acceptable approximation to an ideal EEE interface, and get
\begin{equation}
  \label{eq:opt-b-ideal}
  B^\ast_{\mathrm i} = (1-\rho)\frac{
    \lambda(h^\ast + \TS + \TW) - \rho - \delta
  }{
    \delta \lambda
  }.
\end{equation}

As $B^\ast_{\mathrm i}$ depends on $\rho$, we obtain the load value that makes $B^\ast_{\mathrm i}$
maximum 
and obtain that
\begin{equation}
  \label{eq:r-max-ideal}
  \rho^\ast_{\mathrm i} = \frac{\sqrt{\delta L}}{\sqrt{(h^\ast+\TS+\TW)\mathrm{C} - L}},
\end{equation}
with $\mathrm{C}$ the link capacity and $L$ the average frame size.
Substituting~\eqref{eq:r-max-ideal} into~\eqref{eq:opt-b-ideal} we obtain the
$B$ value that guaranties a given approximation to the ideal sleeping interval
for any load:
\begin{IEEEeqnarray}{rCl}
  \label{eq:opt-b-ideal-bound}
  B^{\max}_{\mathrm i} &=&
  \frac{
    (h^\ast+\TS+\TW)\mathrm{C} + (\delta-1) L
  }
  {
    \delta \mathrm{C}
  }\IEEEnonumber\\
   &&
  -\: \frac{
    2 \sqrt{\delta L} \sqrt{(h^\ast+\TS+\TW)\mathrm{C} - L}
  }
  {
    \delta \mathrm{C}
  }.
\end{IEEEeqnarray}

If we restrict~\eqref{eq:opt-b-ideal-bound} to the usual operating conditions,
then we can neglect the time needed for the transmission of a single frame
($s=L/\mathrm{C} \approx 0$). So, assuming that the hysteresis is at the very
least an order of magnitude higher than the transitions times, we have
\begin{IEEEeqnarray}{rCl}
  \label{eq:opt-b-ideal-bound-aprox}
  \nonumber
  B^{\max}_{\mathrm i} &\approx& \frac{
               h^\ast \mathrm{C} + (\delta -1) L
               }{
               \delta \mathrm{C}
               } - 2\sqrt{
               \frac{
               L h^\ast
               }{
               \delta \mathrm{C}
               } - \frac{
               L^2
               }{
               \delta \mathrm{C}^2
               }
               }\IEEEnonumber\\
             & \approx & \frac{h^\ast}{\delta} + (\delta - 1)\frac{L}{\mathrm{C}} -
               \frac{2}{\sqrt{\delta}}\sqrt{
               \frac{L}{\mathrm{C}}\left(h^\ast - \frac{L}{\mathrm{C}}\right)
               } \IEEEnonumber\\
             & \approx & \frac{h^\ast}{\delta}.
\end{IEEEeqnarray}

\section{Experimental Results}
\label{sec:experimental-results}

In this section we analyze the accuracy of the previous models via simulation.
We focus on the normalized sleeping time instead of the energy consumption for
several reasons. Firstly, it only depends on the LPI governing algorithm and
not on the electrical characteristics of a given NIC\@. While most are
similar, there can be variations among different manufactures. Secondly, NIC
interfaces provide directly to operating system the time spent in each mode
and the number of transitions, but not energy usage estimations. In any case,
as it can be seen from~\eqref{eq:energy-consumption}
and~\eqref{eq:rho-off-definition}, both are interchangeable. For the
experiments we have developed a new EEE simulator with configurable hysteresis
and delay to mimic real hardware. The simulator is available to
download~\cite{HystEEE}. In all simulated scenarios, traffic arrivals follow
either a Poisson or a Pareto process with shape parameter $\alpha=1.8$ to
validate our formulas with self-similar traffic of infinite variance. We
employed constant frame sizes of \SI{1500}{bytes}. Every simulation experiment
has been repeated 20 times with different random seeds and \SI{95}{\percent}
confidence intervals have been calculated. However, considering that the
intervals were very small, we have chosen not to represent them to avoid
cluttering the figures in excess. Finally, the experiments can be divided into
four groups: a first group devoted to assess the accuracy of the delay and
hysteresis models, a second one about the bunching model, a third group that
shows the optimum configuration of the bunching algorithm for a given target
energy expenditure, and the final set of experiments that test the
pre-coalescing algorithm in a hardware test-bed.

\subsection{Model Validation}
\label{sec:model-validation}

In the first set of experiments we validate the energy model for time-based
coalescers with hysteresis developed in Section~\ref{sec:model-time-based}.
\begin{figure}
  \centering
  \includegraphics[width=\columnwidth]{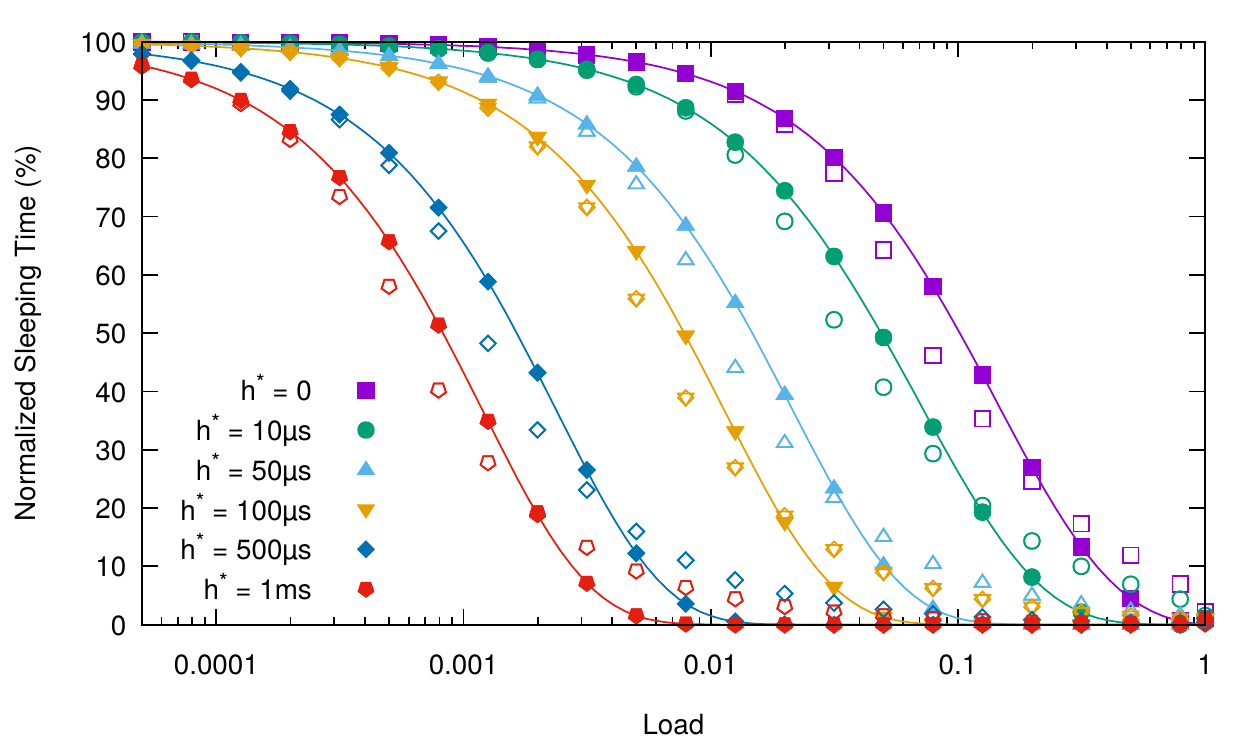}
  \caption{Normalized sleeping time versus load for different hysteresis
    values. Filled points correspond to Poisson traffic, empty points to
    Pareto traffic and continuous lines to model predictions.}
  \label{fig:hyst-model-validation}
\end{figure}
Fig.~\ref{fig:hyst-model-validation} shows the normalized sleeping time
defined in~\eqref{eq:rho-off-definition}, as a hardware-independent proxy for
the energy usage of an EEE interface, from an extremely low load value to an
abnormally high one with a ranging hysteresis time from 0 to \SI{1}{\ms}. In
the figure the filled points correspond to the Poisson traffic, the empty ones
are the results for the Pareto process and the continuous lines are the model
predictions. It comes immediately that the model has enough accuracy and is
able to match the simulation values for every hysteresis and load combination.
Only the results for the self-similar traffic show minor deviations from the
model value. For this kind of traffic, the model overestimates the sleeping
time at low loads, and underestimates it at the highest ones. As expected, we
can also appreciate that there is an inverse relationship between hysteresis
and energy efficiency. From an energy usage point of view it is better to
configure the interface with no hysteresis, though not all commercial
equipment allows to configure small enough hysteresis values.

\begin{figure}
  \centering
  \subfloat[]{\includegraphics[width=\columnwidth]{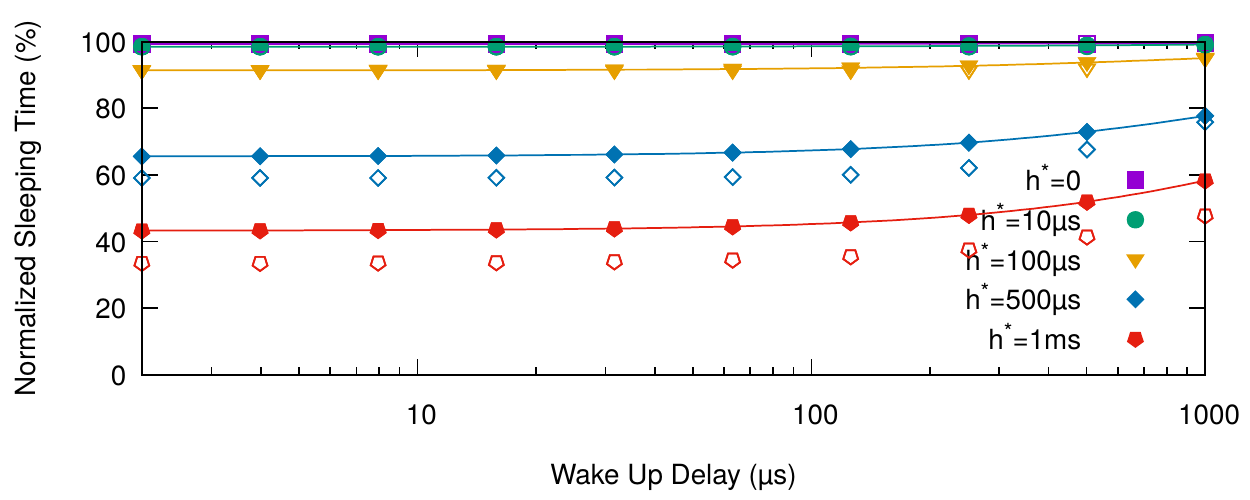}\label{fig:delay-model-validation-0.1}}\hfil
  \subfloat[]{\includegraphics[width=\columnwidth]{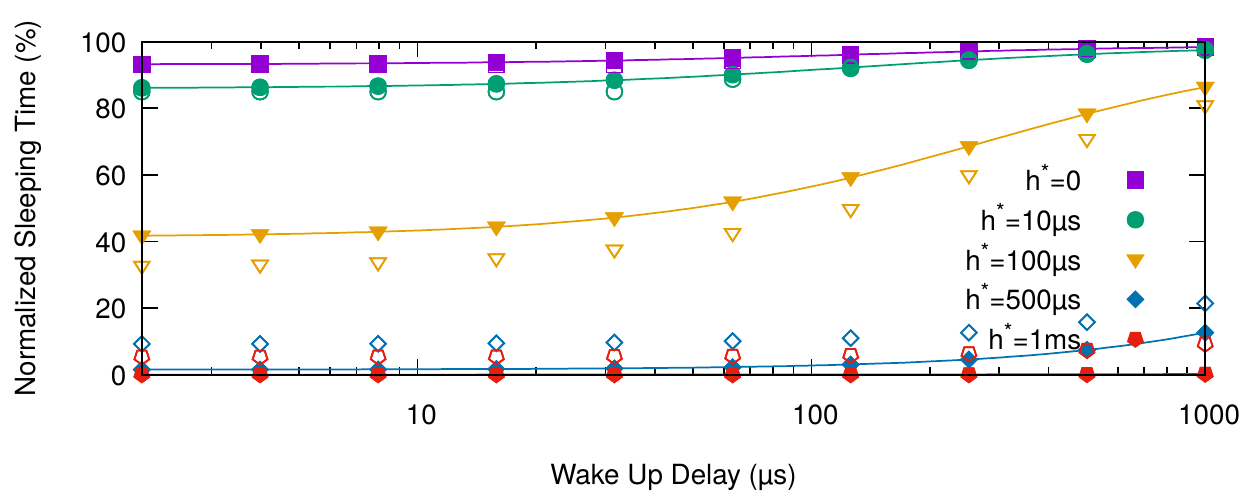}\label{fig:delay-model-validation-1}}
  \hfil\subfloat[]{\includegraphics[width=\columnwidth]{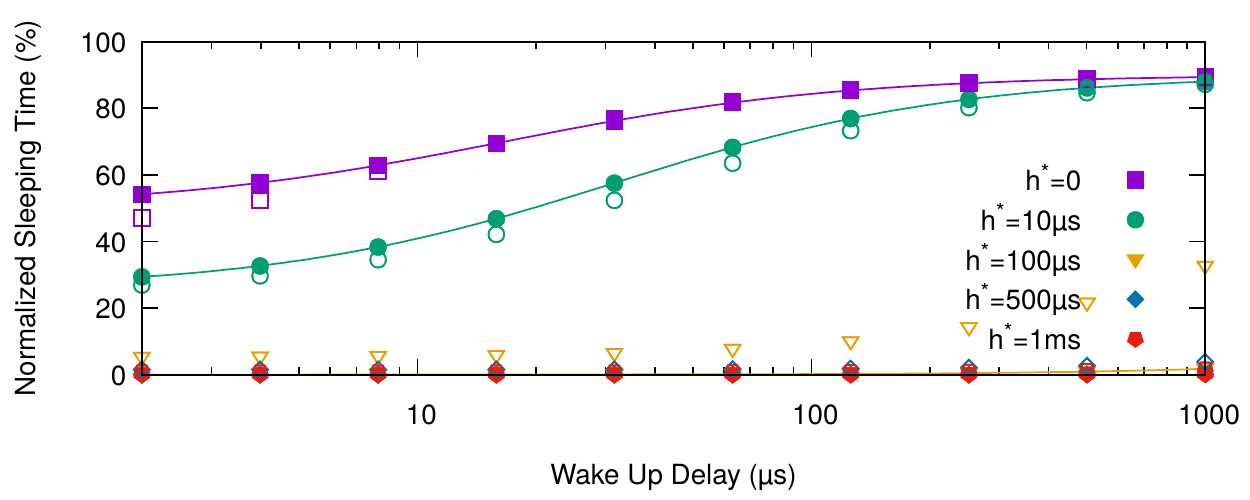}\label{fig:delay-model-validation-10}}\hfil
  \caption{Normalized sleeping time versus wake up delay for different
    hysteresis values and both synthetic Poisson (filled points) and Pareto (empty points)
    traffic of loads \SI{0.1}{\percent}~(a), \SI{1}{\percent}~(b) and
    \SI{10}{\percent}~(c).\label{fig:delay-model-validation}}
\end{figure}
The effect of the wake up delay is represented in
Fig.~\ref{fig:delay-model-validation}. As in the previous experiment, there is
a very good match between the model predictions and the observed values. The
results for the different traffic loads show a similar trend: sleeping time
grows as the wake up delay increases, i.e., the time needed for the interface
to return to active mode since the first frame arrival while idle, augments.
However, for the wake up delay to take significant effect, it needs to be
quite high when compared to the applied hysteresis. For instance, in
Fig.~\ref{fig:delay-model-validation-0.1}, for the $h^\ast=\SI{500}{\us}$ and
$h^\ast=\SI{1}{\ms}$ plots, we only see increasing the sleeping time when the
delay surpasses the \SI{200}{\us} value. In
Fig.~\ref{fig:delay-model-validation-1}, for the $h^\ast=\SI{100}{\us}$ case,
the sleeping time only starts to grow rapidly when the delay surpasses the
\SI{20}{\us} level. Similar conclusions can be drawn from
Fig.~\ref{fig:delay-model-validation-10} for both the no hysteresis and the
$h^\ast=\SI{10}{\us}$ cases. In every case we see that the hysteresis value
has a much greater effect than the delay on energy consumption.

\begin{figure}
  \centering
  \includegraphics[width=\columnwidth]{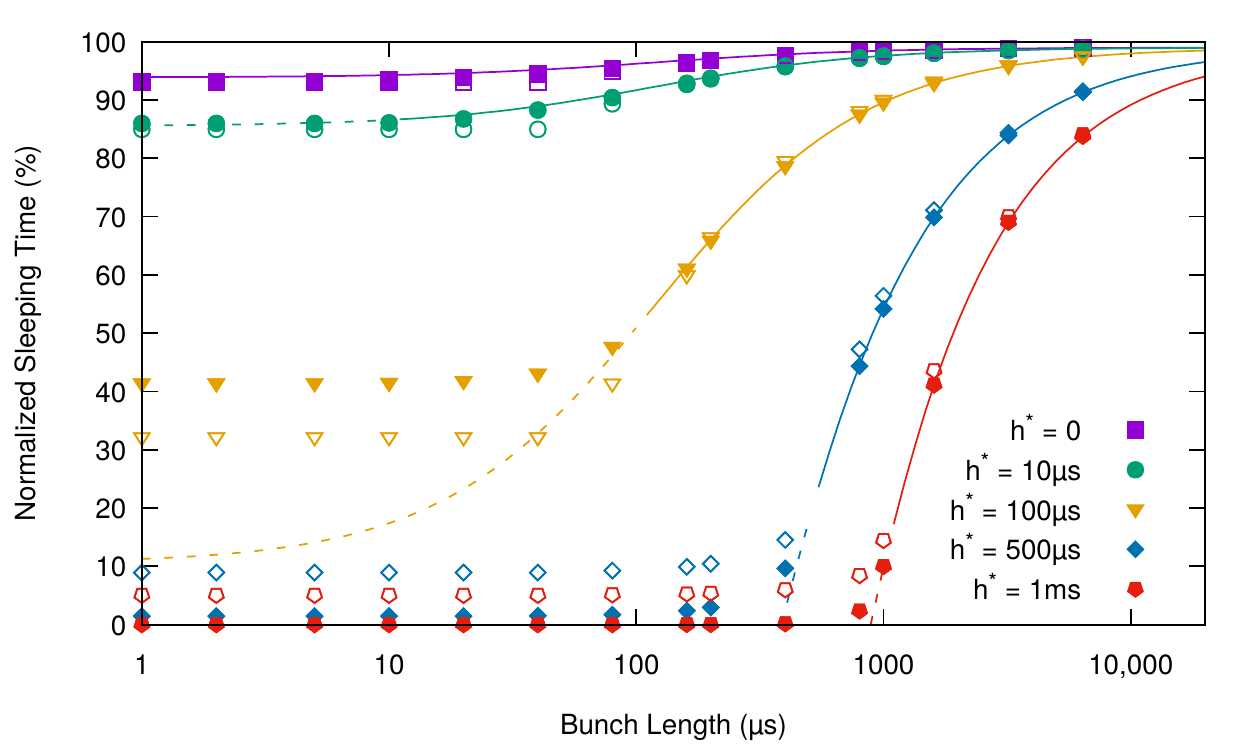}
  \caption{Normalized sleeping time versus bunch length for a 1\% load and
    various hysteresis lengths with Poisson and self-similar traffic. Dashed 
    lines represent values out of the model valid
    range.\label{fig:bunching-no-delay}}
\end{figure}

\subsection{Pre-coalescing Algorithm Evaluation}
\label{sec:pre-coal-algor-1}

The next set of experiments focuses on validating the model for the
pre-coalescing algorithm as in~\eqref{eq:rlpi-single-hyst}. To this end, we
have fed both Poisson and self-similar Pareto traffic to a frame coalescer
that implements our bunching algorithm. The software used has been released
with an open-source license, and it is available for
download~\cite{EEEBunch}.

Fig.~\ref{fig:bunching-no-delay} compares the model predictions to the results
of applying an increasing bunching length to an EEE interface configured with
different hysteresis values and \SI{1}{\percent} load,\footnote{We have
  experimented with different load values (from \SIrange{0.1}{50}{\percent}),
  obtaining comparable results. We have omitted them in the paper for the sake
  of brevity.} albeit without wake up delay. As in the previous experiments,
the simulated results are shown with points and the model predictions with
lines. To account for the fact that the model is only valid for bunching
lengths greater than the interface hysteresis, we have employed dashed lines
for those regions where the model is not expected to hold. All in all, the
results confirm the accuracy of the model within its valid region. We can also
appreciate how the bunching technique improves the energy efficiency of the
interface, even if at the cost of increased delay. As a rule of thumb, we see
that we need a bunch length, i.e., the maximum delay suffered by a frame, ten
times the duration of the hysteresis for significant savings. On the other
hand, for low bunching lengths, the model underestimates the obtained savings.
This can be seen very clearly in the $h^\ast=\SI{100}{\us}$ results.

\begin{figure}
  \centering
  \includegraphics[width=\columnwidth]{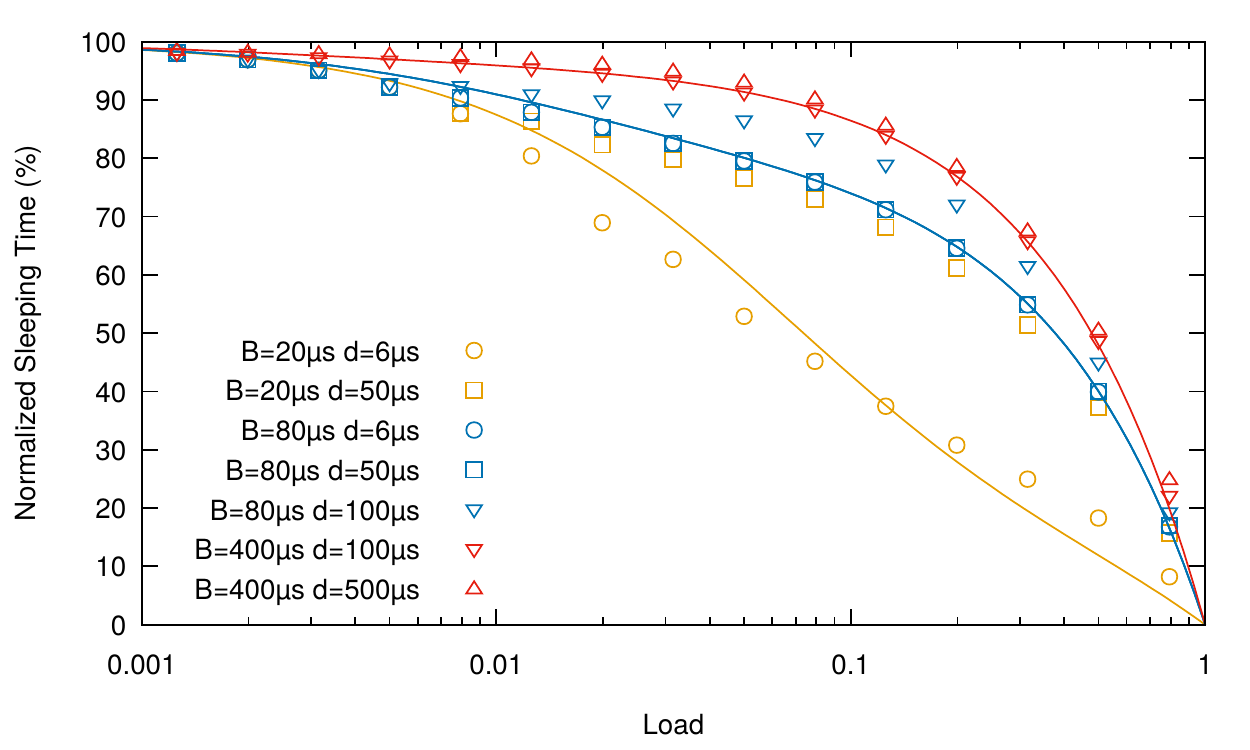}
  \caption{Normalized sleeping time versus load for different bunch length
    and delay combinations with self-similar traffic. Hysteresis is set to
    \SI{10}{\us}. The continuous line represents the
    model~\eqref{eq:rlpi-single-hyst}.}
    \label{fig:bunching-model-delay}
\end{figure}

In Fig.~\ref{fig:bunching-model-delay} we represent the evolution of the
sleeping time for different Pareto traffic loads and a set of different
combinations of bunching length and wake up delay.\footnote{We have obtained
  similar results with Poisson traffic, but they have not been plotted to
  avoid cluttering the figure in excess.} The hysteresis is otherwise fixed to
a relatively low value of \SI{10}{\us} so that delay effects are more evident.
Recall that for the wake up delay to be noticeable, it has to be greater than
the hysteresis, as discussed previously when presenting
Fig.~\ref{fig:delay-model-validation}. It is clear again that the
model~\eqref{eq:rlpi-single-hyst} is a good predictor for the experimental
results, at least as long as it stays in its valid region, that is, when the
wake up delay is shorter than the bunching length ($d < B$). As expected, the
higher the wake up time, the longer the interface stays in the LPI mode. We
can also observe that it is not worth to use delay at the network interface if
it is not greater than the bunching length. 

\subsection{Pre-coalescing Approximation to Frame Transmission and Ideal
  Energy Savings}
\label{sec:pre-coal-tuning}

\begin{figure}
  \centering
  \subfloat[]{\includegraphics[width=\columnwidth]{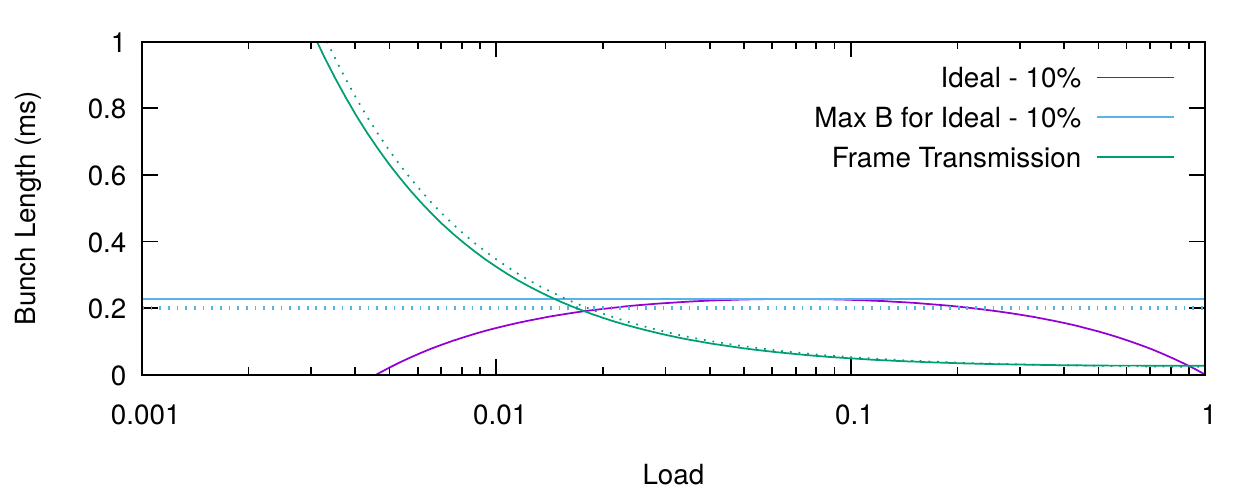}\label{fig:bunch-lengh-opt-aggresive}}\hfil
  \subfloat[]{\includegraphics[width=\columnwidth]{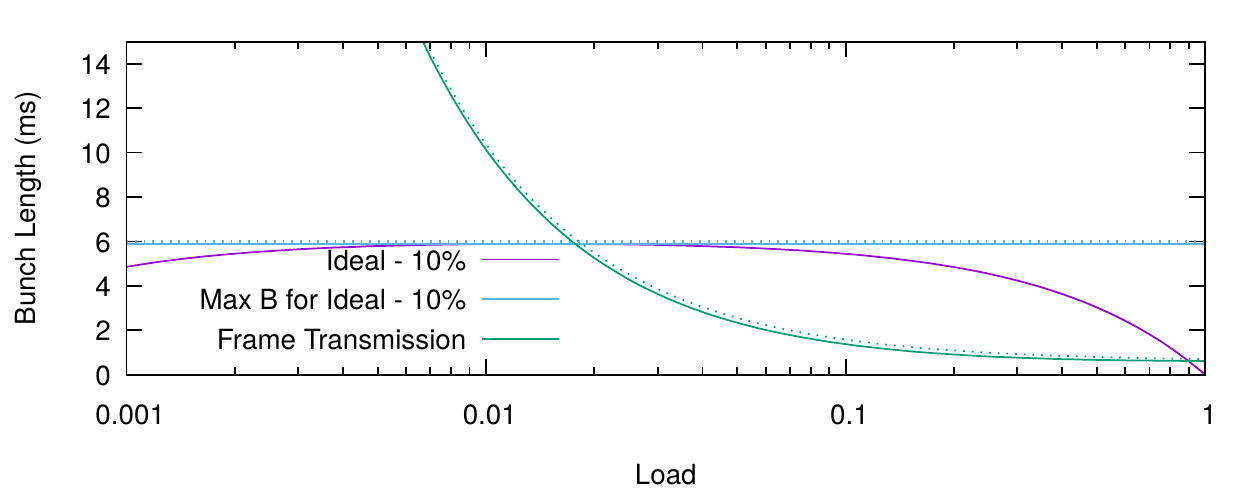}\label{fig:bunch-lengh-opt-conservative}}
  \caption{Bunch length needed to obtain a given target consumption for
    varying Pareto traffic loads and (a)~aggressive and (b)~non~aggressive
    settings.\label{fig:bunch-length-opt}}
\end{figure}
This set of experiments explores the tuning of the bunching length at the
pre-coalescer in order for the NIC to obtain results comparable to those of:
\begin{enumerate*}[a)]
\item an ideal EEE interface, i.e., one that only stays awake while transmitting
  traffic and in LPI otherwise, with a \SI{10}{\percent} error margin; 
\item and to
  those of an interface with no hysteresis nor delay that employs the frame
  transmission algorithm.
\end{enumerate*}
We consider two different configurations for the
interface, with different delay and hysteresis parameters that we have called
\emph{aggressive} and \emph{non aggressive} settings. The actual parameters
have been selected in accordance with the characteristics of current actual
hardware~\cite{cisco7000,dell1524p} and are the ones shown in
Table~\ref{tab:switch-parameters}.

Fig.~\ref{fig:bunch-length-opt} shows the necessary bunching length to obtain
energy savings in line with those of frame transmission and those of an ideal
interface for varying traffic load conditions. At the same time, it shows the
accuracy of the simplifications carried out in
Section~\ref{sec:achi-frame-transm}. The exact results are shown with
continuous lines, while the approximated values with dotted lines. In any
case, we can conclude that the approximations provide good enough results. It
is also clear from both plots that~\eqref{eq:opt-b-ideal-bound}, and
consequently~\eqref{eq:opt-b-ideal-bound-aprox}, permit to configure the
bunching length without regards to the actual traffic load.

We can also deduce some additional conclusions. At the highest loads, the
necessary bunching length diminishes both for the frame transmission target and
for the approximation to the ideal interface. This is because at the highest
loads the energy savings are always lowest. This is more pronounced in frame
transmission, as it obtains modest savings from moderate loads onward. On the
contrary, for the lowest loads it is almost impossible to get the same results
as frame transmission does, as it is able to obtain almost ideal energy savings. 
However, if we use the ideal conditions with a \SI{10}{\percent} error margin 
as a reference, we see
that, for the aggressive settings (Fig.~\ref{fig:bunch-lengh-opt-aggresive}),
there is no need to employ pre-coalescing. On the contrary, for the
conservative settings (Fig.~\ref{fig:bunch-lengh-opt-conservative}) and the
represented loads, bunching is always needed. Obviously, if we further
increased the load ranges to include even lower values, we would be able to
find loads were bunching would not be needed. Please note, however, that the
considered range of loads is already big enough to cover for all usual cases.

\begin{figure}
  \centering
  \includegraphics[width=\columnwidth]{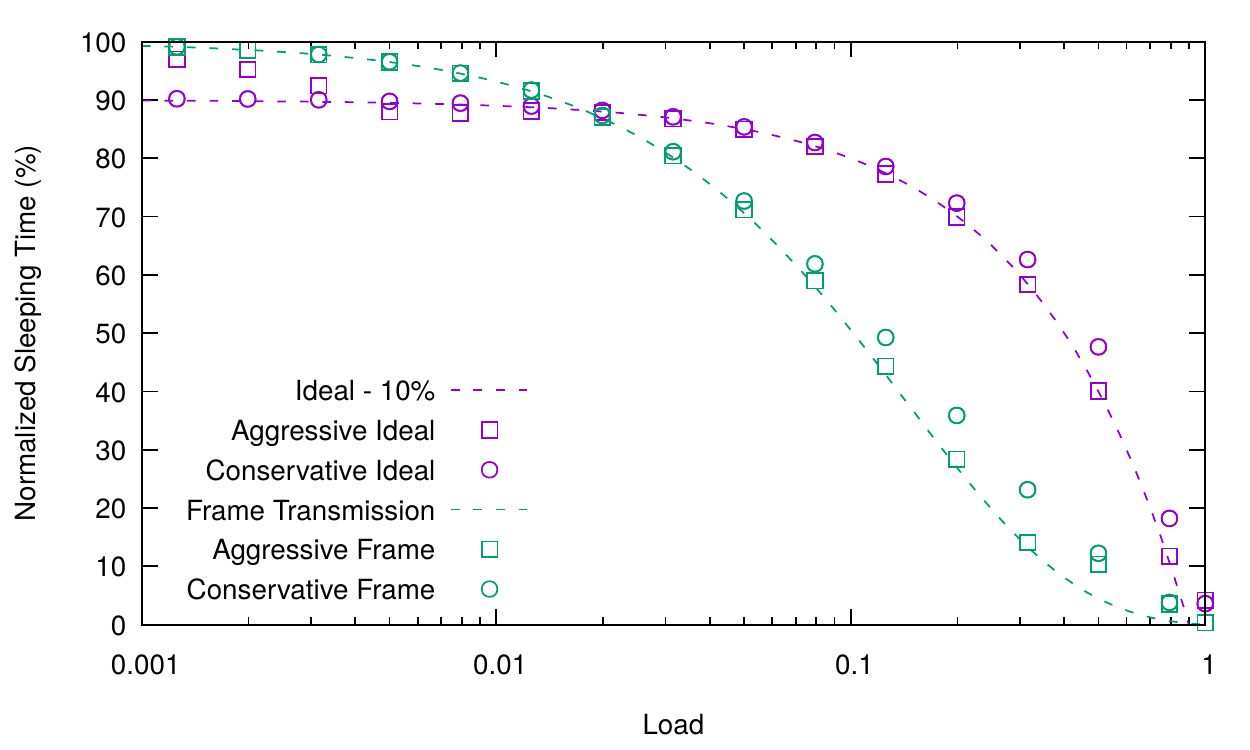}
  \caption{Normalized sleeping time when using the optimum bunch length to get
    a target consumption.}
  \label{fig:opt-bunch-results}
\end{figure}
We have represented in Fig.~\ref{fig:opt-bunch-results} the normalized
sleeping time obtained when using the approximated bunching lengths of
Fig.~\ref{fig:bunch-length-opt}. We can confirm that the calculated bunching
lengths are enough to obtain the desired energy savings. In fact, for moderate
to high loads, the employed bunching lengths slightly outperform their target
value. The same happens for low loads and the ideal settings. In this case it
is because no bunching is necessary to get or even improve the target of less
than \SI{10}{\percent} difference when compared to the ideal algorithm.


\begin{figure}
  \centering
  \subfloat[]{\includegraphics[width=\columnwidth]{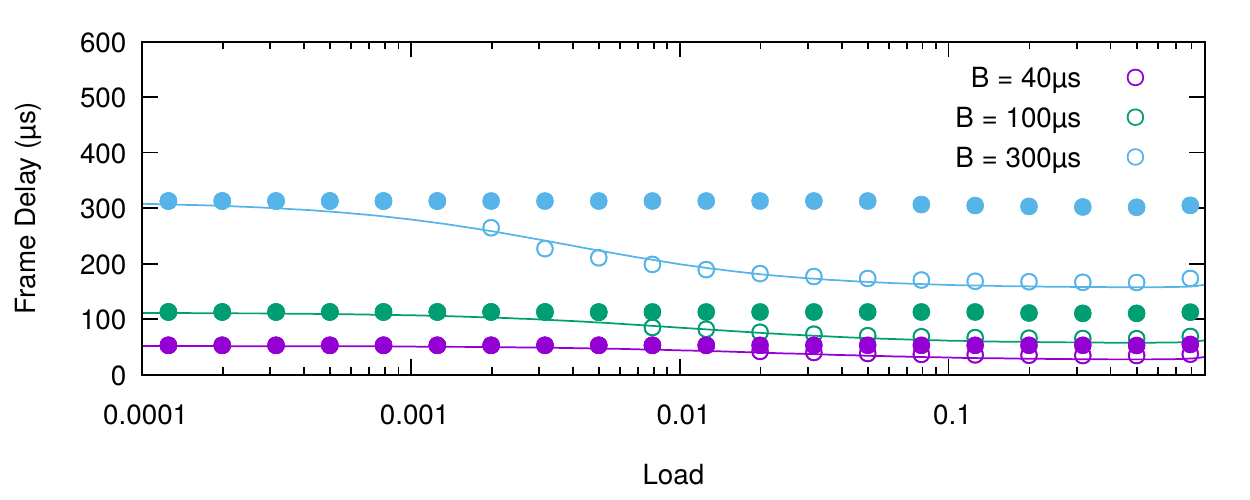}\label{fig:hist-delay-bunch-h20}}\hfil
  \subfloat[]{\includegraphics[width=\columnwidth]{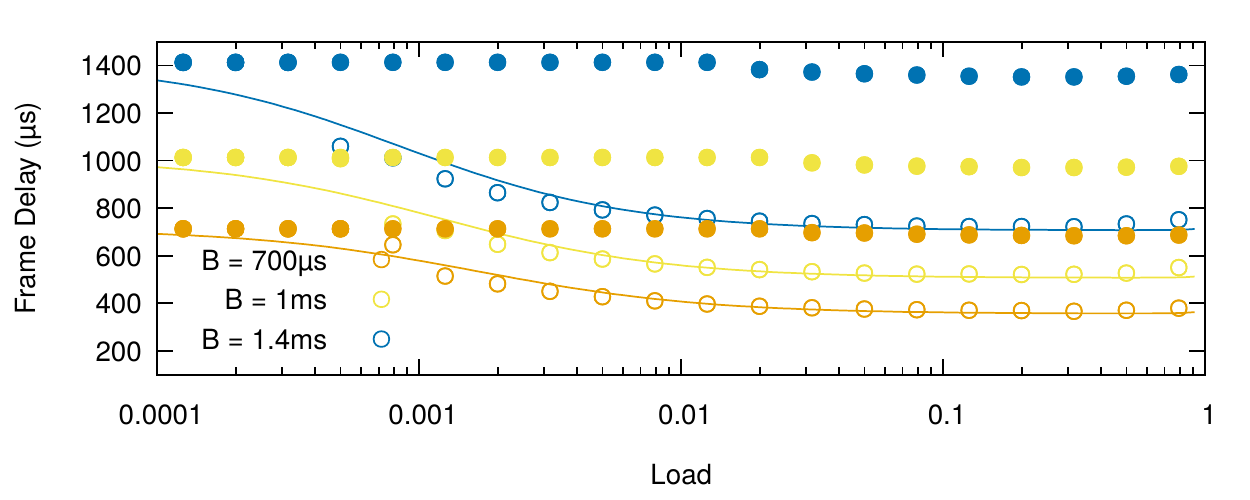}\label{fig:hist-delay-bunch-h600}}
  \caption{Average frame delay for different traffic loads when using the
    pre-coalescing algorithm (empty circles). Filled circles show the
    \SI{95}{\percent} percentile of frame delay: (a)~$h^\ast=\SI{20}{\us}$ 
    and (b)~$h^\ast=\SI{600}{\us}$.}
  \label{fig:hist-delay-bunch}
\end{figure}
In Fig.~\ref{fig:hist-delay-bunch} we show the effects on average frame delay
of some of the recommended bunching lengths shown in
Fig.~\ref{fig:bunch-length-opt}. We have shown both the average frame delay
and its \SI{95}{\percent} percentile as empty and filled dots, respectively.
The expected average values from the
model~\eqref{eq:av-delay-marshall-poisson} are represented with continuous
lines. Results show that the model is a good predictor for the average value
and that the delay is dominated by the pre-coalescing process. 
It is important to realize that, given that the
bunching process dominates the delay, the worst case per-frame delay is very
stable and can be approximated by the $B$ parameter.

\subsection{Actual Hardware Results}
\label{sec:actu-hardw-results}

In our last experiment we measured the performance of the pre-coalescing
technique with two \SI{10e9}{\bit\per\s} NICs. In particular we used two
Intel~X550 controllers~\cite{intel_ether_contr_x550_datas} connected to an
Intel~Core~i7-3770 computer running Ubuntu Linux~19.04. Sadly, the hardware
does not report the accumulated time spent in LPI, only the number of times
the interface enters LPI mode.\footnote{As a matter of fact, the official
  driver does not yet support EEE~\cite{intel_x550_features}. We have made
  available a simple tool to configure and query EEE mode in x550 NICS at
  \url{https://migrax.github.io/eee-X550/}.}

We fed the NIC with self-similar traffic traces with packet sizes of 1500
bytes and different average rates. Then we captured the traffic received at
the other end. With the number of LPI transitions and the gaps between the
frames in the received trace, we have calculated the average time spent in
LPI\@. Although this device allows to completely disable hysteresis, we have
used $h^\ast=\SI{20}{\us}$ to account for the most advantageous value
available from other vendors. A bunching length of $\SI{200}{\us}$ was chosen
according to~\eqref{eq:opt-b-ideal-bound-aprox}. The results are shown in
Table~\ref{tab:x550-results}.
\begin{table*}
  \centering
  \caption{Time Spent in LPI Mode on an Intel x550 Interface with
    $h^\ast=\SI{20}{\us}$}
  \label{tab:x550-results}
  \begin{tabu}{rSSSSSS}\\\toprule
    &\multicolumn{6}{c}{No Pre-coalescing / \SI{200}{\us}
      Pre-coalescing}\\\cmidrule{2-7}
    \multicolumn{1}{c}{Rate} & \multicolumn{2}{c}{LPI events$/$s} &
    \multicolumn{2}{c}{Av. Duration} & \multicolumn{2}{c}{Time in LPI (\%)}\\\midrule
    \SI{10e6}{\bit\per\s}
    & 803 & 804 &
    \SI{1.21}{\ms}&\SI{1.21}{\ms} &
    97.8 & 97.9 \\
    
    \SI{100e6}{\bit\per\s}
    & 7977 & 3004
    & \SI{98.6}{\us} & \SI{302}{\us}
    & 78.6 & 90.6\\
    
    \SI{850e6}{\bit\per\s}
    &5334 & 2664
    & \SI{107}{\us} & \SI{260}{\us}
    & 57.3 & 69.1 \\
    \bottomrule
  \end{tabu}
\end{table*}
For very low rates, the LPI event rate only depends on the frame interarrival
times (about \SI{1.2}{\ms} for \SI{10e6}{\bit\per\second}). As the rate
increases the pre-coalescer starts grouping individual arrivals. For the
\SI{100e6}{\bit\per\second} trace, frames arrive, on average, every
\SI{120}{\us} without pre-coalescing while, with pre-coalescing, a bunch of an
average of two frames arrives every
$\SI{120}{\us} + \SI{200}{\us} = \SI{320}{\us}$. This results in almost $2.6$
more transitions per second for the non pre-coalescing setup, yielding to
lower energy savings due to the time spent in hysteresis and in transitions
for each LPI event. Finally, for higher rate traces, several frames are served
in the same busy period for both scenarios. This causes the event rate to lose
direct dependence on individual frame interarrival times. Results show that,
as expected, for a low hysteresis value, no pre-coalescing is necessary for
very low average rates. However, as the rate increases, even for modest loads,
pre-coalescing is able to significantly increase energy savings.



\section{Conclusions}
\label{sec:conclusions}

This paper presents a new model for EEE interfaces with hysteresis, and shows
that many existing EEE networking devices are unable to obtain significant
energy savings with their available configuration. To improve their energy
efficiency, the paper proposes to implement frame coalescing before traffic
reaches the interface, and provides an analytic model for this technique. This
model allows us to derive the optimum coalescing parameters to obtain a
desired target energy consumption at the EEE device, when the configuration
parameters at the device are known in advance.

The study includes four groups of experiments. The first set demonstrates that
the energy model considering hysteresis and delay has enough accuracy and is
able to match the simulation values for every relevant hysteresis and load
combination. From an energy usage point of view, it is always best to
configure the interface with no hysteresis. Furthermore, the wake up delay
needs to be quite high to compensate for the negative effects of hysteresis.
The second group of experiments focus on validating the model for the bunching
algorithm as in~\eqref{eq:rlpi-single-hyst}. The results confirm the accuracy
of the model within its valid region (for bunching lengths longer than the
interface hysteresis). Besides, we could appreciate how the bunching technique
improves the energy efficiency of the interface, albeit at the cost of an
increased delay. We can also see that it is not worth to use delay at the
interface if it is not greater than the bunching length. The experiments also
analyzed the ideal tuning of the bunching length to mimic frame transmission
performance or to get close to ideal energy usage. The proper tuning results
in a bunch length directly proportional to the hysteresis value, and inversely
proportional to the maximum deviation from the ideal energy consumption. This
bunch interval has direct consequences on frame delay. For the worst case and
for almost any traffic load, the maximum frame delay is very close to the
bunch interval. So, for delay-stringent applications, it is important to
either employ interfaces with low hysteresis values, or to compromise energy
savings allowing greater deviations from the ideal energy consumption.
Finally, the fourth set of experiments validated the results on a real
hardware test-bed equipped with EEE capable NICs.

The pre-coalescing technique is a valid approach to overcome the very high
hysteresis values available at most EEE equipment. It has the additional
advantage that it can take into account traffic needs, in contrast with using
delay at the interface, which has to be the same for all the transit traffic.
In any case, EEE equipment should provide the option to select lower
hysteresis values or even no hysteresis at all to the network administrator.
This is the best way to get significant energy savings at relevant traffic
loads without excessive frame delay.

\bibliography{bunching}
\bibliographystyle{unsrtnat}

\end{document}